\documentclass{ws-ijmpcs}

\begin{document}

\markboth{Yu-Feng Li} {Overview of the Jiangmen Underground Neutrino
Observatory}

%
\catchline{}{}{}{}{}
%

\title{Overview of the Jiangmen Underground Neutrino Observatory (JUNO)}

\author{Yu-Feng Li}

\address{Institute of High Energy Physics, Chinese Academy of Sciences,
Beijing 100049, China\\
{\textrm{liyufeng@ihep.ac.cn}}}

\maketitle

\begin{history}
\received{Day Month Year}
\revised{Day Month Year}
\end{history}

\begin{abstract}
The medium baseline reactor antineutrino experiment, Jiangmen
Underground Neutrino Observatory (JUNO), which is being planed to be
built at Jiangmen in South China, can determine the neutrino mass
hierarchy and improve the precision of three oscillation parameters
by one order of magnitude. The sensitivity potential on these
measurements is reviewed and design concepts of the central detector
are illustrated. Finally, we emphasize on the technical challenges
we meet and the corresponding R$\&$D efforts. 
\end{abstract}

\ccode{PACS numbers: 14.60.Pq, 29.40.Mc, 28.50.Hw, 13.15.+g}

\section{Introduction}
After the discovery of non-zero $\theta_{13}$ in latest reactor
\cite{DYB,DYBcpc,DYBspectra,DC,RENO} and accelerator
\cite{T2K,MINOS} neutrino oscillation experiments, the neutrino mass
hierarchy (i.e., the sign of $\Delta m^2_{31}$ or $\Delta m^2_{32}$)
and lepton CP violation are the remaining oscillation parameters to
be measured in the near future. The methods of determining the
neutrino mass hierarchy (MH) include the matter-induced oscillations
in the long-baseline accelerator neutrino experiments
\cite{HK,LBNE,LBNO} and atmospheric neutrino experiments
\cite{INO,PINGU}, and the vacuum oscillations in the medium baseline
reactor antineutrino experiments \cite{DYB2t,DYB2e,JUNO,RENO50}.

\begin{figure}
\begin{center}
\begin{tabular}{c}
\includegraphics*[bb=20 20 591 341, width=0.7\textwidth]{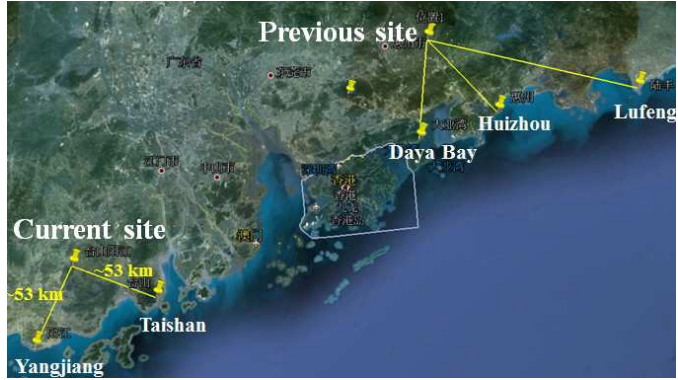}
\end{tabular}
\end{center}
\caption{Layout of the JUNO experimental design. Current site makes
use of the Taishan and Yangjiang reactor complexes, where the
previous site is not considered because the third reactor complex
(Lufeng) is being planed. \label{f1}}
\end{figure}

The Jiangmen Underground Neutrino Observatory (JUNO) is a
multipurpose liquid scintillator (LS) neutrino experiment, whose
primary goal \cite{JUNO} is to determine the neutrino mass hierarchy
using reactor antineutrino oscillations. The layout of JUNO is shown
in Fig. \ref{f1}, where the candidate site is located at Jiangmen in
South China, and 53 km away from the Taishan and Yangjiang reactor
complexes. The overburden for the experimental hall is required to
be larger than 700 meters in order to reduce the muon-induced
backgrounds.

\section{Physics potential}
Because the relative size of two fast oscillation components is
different ($|\Delta m^2_{31}|>|\Delta m^2_{32}|$ or $|\Delta
m^2_{31}|<|\Delta m^2_{32}|$), the interference between the two
oscillation frequencies in the reactor antineutrino energy spectrum
gives us discrimination ability of two different MHs (normal or
inverted). The discrimination power is maximized when the $\Delta
m^2_{21}$ oscillation is maximal (see Figure 1 of Ref. [15]).

\begin{figure}
\begin{center}
\begin{tabular}{c}
\includegraphics*[bb=28 22 292 222, width=0.65\textwidth]{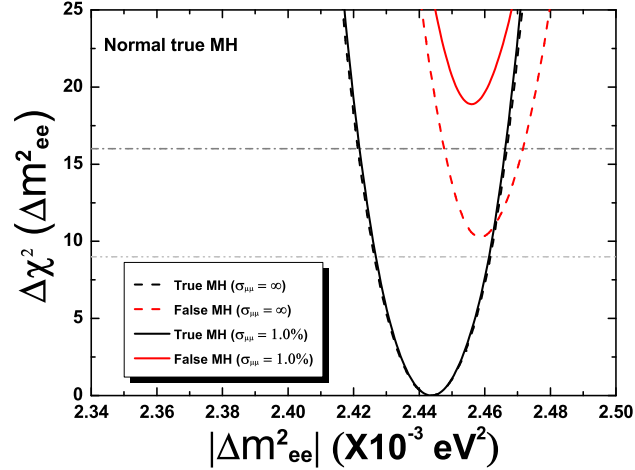}
\end{tabular}
\end{center}
\caption{MH sensitivity of JUNO using reactor antineutrino
oscillations. The vertical difference of the curves for the true and
false MHs is the discriminator defined in Eq. (1). The solid and
dashed lines are for the analyses with and without the prior
measurement of $\Delta m^2_{\mu\mu}$. \label{f2}}
\end{figure}

To calculate the sensitivity of MH determination at JUNO, we assume
the following nominal setups. A LS detector of 20 kton is placed 53
km away from the Taishan and Yangjian reactor complexes. The
detailed distance and power distribution of reactor cores summarized
in Table 1 of Ref. [15] is used to include the reduction effect of
baseline difference. In the simulation, we use nominal running time
of six years, 300 effective days per year, and a detector energy
resolution $3\%/\sqrt{E{\rm (MeV)}}$ as a benchmark. A normal MH is
assumed to be the true one while the conclusion won't be changed for
the other assumption. The relevant oscillation parameters are taken
from the latest global analysis \cite{global1}. To illustrate the
effect of energy non-linearity and the power of self-calibration, we
assume a residual non-linearity curve parametrized in Figure 3 of
Ref. [15] and a testing polynomial non-linearity function with
$100\%$ uncertainties for the coefficients. Taking into account all
above factors in the least squares method, we can get the MH
sensitivity as shown in Fig. \ref{f2}, where the discriminator is
defined as
\begin{equation}
\Delta \chi^2_{\text{MH}}=|\chi^2_{\rm min}(\rm Normal)-\chi^2_{\rm
min}(\rm Inverted)|\,,
\end{equation}
and $\Delta m^2_{ee}$ and $\Delta m^2_{\mu\mu}$ are the effective
mass-squared differences \cite{Parke} in the electron and muon
neutrino disappearance experiments, respectively. From the figure we
can learn that a confidence level of $\Delta \chi^2 \simeq 11$ is
achieved for the reactor-only analysis, and it will increase to
$\Delta \chi^2 \simeq 19$ by using a prior measurement of $\Delta
m^2_{\mu\mu}$ ($1\%$).

Other important goals of JUNO include the precision measurement of
oscillation parameters and unitarity test, observation of supernova
neutrinos, geo-neutrinos, solar neutrinos and atmospheric neutrinos,
and so on. Using reactor antineutrino oscillations, we can measure
three of the oscillation parameters (i.e. $\sin^2\theta_{12}$,
$\Delta m^2_{21}$ and $|\Delta m^2_{31}|$) better than $1\%$.

\section{Design concepts}
\begin{figure}
\begin{center}
\begin{tabular}{c}
\includegraphics*[bb=24 20 596 400, width=0.7\textwidth]{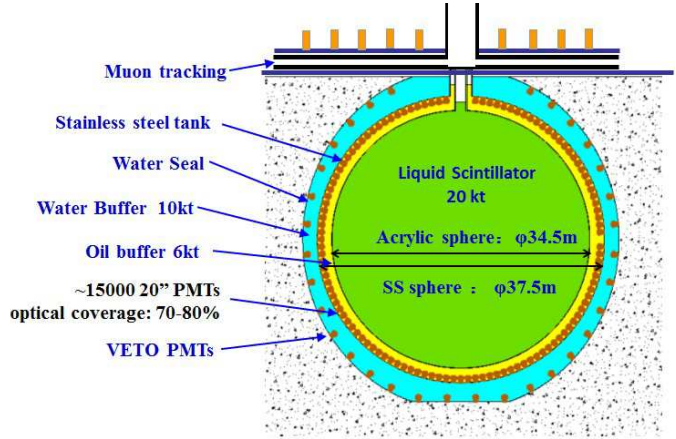}
\end{tabular}
\end{center}
\caption{A basic design of the JUNO central detector (see the
context for details). \label{f3}}
\end{figure}

The design of the central detectors is still open for different
options. One basic option is shown in Fig. \ref{f3}, where the
concept of three separated layers is used for better radioactivity
protection and muon tagging. The inner acylic tank contains 20 kton
linear alkylbenzene (LAB) based LS as the antineutrino targets.
15,000 20-inch photomultiplier tubes (PMTs) are installed in the
internal surface of the outer stainless steel tank. 6 kton mineral
oil is filled between the inner and outer tanks as buffer of
radioactivities. 10 kton high-purity water is filled outside the
stainless steel tank. It serves as a water Cherenkov detector after
being mounted with PMTs. Other design concept contains the balloon
option, single tank option, PMTs module option and mixtures among
them. The energy resolution, radioactivity level and technical
challenges are the main concerns of different options.

To obtain an unprecedent energy resolution level of $3\%/\sqrt{E{\rm
(MeV)}}$ (or 1,200 photon electrons per MeV) is a big challenge for
a LS detector of 20 kton. Much better performance for PMTs and LS is
required. R$\&$D efforts to overcome the above challenges are being
developed within the JUNO working groups. A new type \cite{MCP} of low-cost
high-efficiency PMTs is being designed, which uses the micro channel
plate (MCP) as the dynode and receives both the transmission and
reflection light using the reflection photocathode. A coverage level
of $80\%$ can be realized with a careful consideration of the PMTs
spacing and arrangement. Moreover, highly transparent LS with longer
attenuation length ($>$30 m) is also being developed. Both the
method of LS purification by using ${\rm Al}_2{\rm O}_3$ and the
distillation facility and the method to increase the light yield are
considered.

\section{Conclusion}
JUNO is designed to determine the neutrino MH ($3\sigma-4\sigma$ for
six years) and measure three of the oscillation parameters better
than $1\%$ using reactor antineutrino oscillations. It can also
detect the neutrino sources from astrophysics and geophysics. It has
strong physics potential, meanwhile contains significant technical
challenges. This program is supported from the Chinese Academy of
Sciences and planed to be in operation in 2020.

%

\end{document}